\documentclass[a4paper,11pt]{article}

\usepackage{jcappub}
\usepackage{subcaption}

\title{A Radical Solution to the Hubble Tension Problem}

\author{Timothy Clifton}
\author{and Neil Hyatt}

\affiliation{Department of Physics and Astronomy, Queen Mary University of London, UK}

\emailAdd{t.clifton@qmul.ac.uk}

\abstract{
The Hubble tension has proven to be stubbornly persistent, despite widespread efforts to relax it. As a possible resolution of this problem we propose a radical alternative to the way in which cosmological models are viewed. Specifically, we consider building cosmological models from spaces that exhibit intrinsic symmetries, rather than as space-times with explicit symmetry. This change in perspective allows statistical homogeneity and isotropy to be maintained, while relaxing some strong mathematical constraints that the standard approach imposes. We show that a Hubble tension arises naturally in our new approach, and that (as a corollary) a prediction can be made for the radial component of the Baryon Acoustic Oscillations. Our prediction appears to be  consistent with the DESI first-year data release, which has otherwise been interpreted as evidence for dynamical dark energy.
}

\notoc

\begin{document}
\maketitle
\flushbottom

\newpage

\section{Introduction}\label{sec:introduction}

The Hubble tension is a mismatch between the inferred value of the Hubble constant, $H_0$, when interpreting data from astronomical sources in the early and late Universes \cite{many}. This is most directly seen when comparing the value $H_0=67.27 \pm 0.60 \, {\rm km} \, {\rm s}^{-1} \, {\rm Mpc}^{-1}$ from  Planck's observations of the Cosmic Microwave Background (CMB) \cite{cmb}, with the value of $H_0=73.04 \pm 1.04 \, {\rm km} \, {\rm s}^{-1} \, {\rm Mpc}^{-1}$ from the SHOES observations of relatively nearby supernovae \cite{shoes}. Other observations of the early and late Universe seem to neatly line up with the two values quoted above, providing a picture in which measurements of $H_0$ made within these two different regimes seem to be discordant with each other. Taken at face value, this indicates a very serious problem with our understanding of the Universe. Even worse, it seems to be a problem that is strongly resistant to resolution \cite{review}.

We wish to propose a possible explanation for this tension that is based on removing mathematical restrictions that often appear in cosmological modelling, but which may not be necessary, and which might even be inappropriate for modelling the real Universe. In particular, we consider the following two foundational assumptions, which are involved in most attempts at constructing models of our statistically homogeneous and istropic Universe:
\begin{itemize}
\item[(1)] At some suitably chosen time, spatial averages of quantities such as expansion and spatial curvature can be used to represent the state of the Universe.
\item[(2)] The Universe can be represented by a homogeneous and istotropic space-time with expansion and spatial curvature corresponding to those spatial averages.
\end{itemize}
At a superficial level, these two statements seem very similar. We will show that this is not the case, and that assumption (2) is considerably stronger and more restrictive than assumption (1). We will argue that the Hubble tension may be a manifestation of this additional restriction, and that by discarding assumption (2) we can create a more general class of cosmological models that may be able to account for the Hubble tension.

The basic idea we are proposing was discussed some time ago by Ellis and Stoeger, who cast cosmological inferences from astronomical observations as a ``fitting problem'' \cite{fitting}. Our approach to investigating the fitting problem in the context of the Hubble tension will make extensive use of Buchert's spatial averaging formalism \cite{buchert1, buchert2}, as well as R$\ddot{\rm a}$s$\ddot{\rm a}$nen's treatment of the optical properties of statistically homogeneous and isotropic cosmologies \cite{ras1, ras2}. We will also coopt the language of Collins and Szafron in their work on cosmologies with ``intrinsic symmetries'' \cite{col1, col2, col3}, which make clear the distinction between a symmetry that exists on a 3-space (an intrinsic symmetry) and a symmetry that exists on a 4-dimensional space-time (an explicit symmetry). Our aim will be to use cosmological models with intrinsically homogeneous and isotropic 3-spaces, in a fitting procedure of the type envisaged by Ellis and Stoeger, in order to interpret cosmological observables {\it without} assuming the existence of a cosmological space-time with explicit homogeneity and isotropy.

The plan of this paper is as follows: in Section \ref{sec:modelling} we give an overview of Buchert's approach to spatial averaging, and show that in general there exist strong constraints on space-times with explicit symmetries that share the properties of these averages. We take this to indicate that building cosmologies from solutions to Einstein's equations with explicit symmetries may not be appropriate for modelling the real Universe. We then introduce R$\ddot{\rm a}$s$\ddot{\rm a}$nen's approach to calculating the optical properties of statistically homogeneous and isotropic universes in Section \ref{sec:optics}, and discuss how an observer in such a universe could use a Friedmann-Robertson-Walker (FRW) cosmology as a fitting model. In Section \ref{sec:results} we show that the use of such fitting models can lead directly to a Hubble tension, in keeping with the discussion in Ref. \cite{ten3}. We also investigate the possibility that our hypothesis for relaxing the Hubble tension could be checked by analysing the radial component of the Baryon Acoustic Oscillations, and suggest that the DESI first-year data release may show hints of the required signature. We then conclude in Section \ref{sec:disc}. Appendices A and B contain mathematical proofs of statements used in our arguments, and Appendix C contains a table summarizing notation.

\section{Spatial averaging and cosmological modelling} \label{sec:modelling}

In this section we consider the scalar averaging formalism prescribed by Buchert \cite{buchert1, buchert2}, which provides a simple and mathematically elegant way of deriving spatial averages on a foliation of space-time that is orthogonal to the flow of matter. This approach has been much studied since its publication, and is perhaps the most widely known and successful approach to the problem of extracting large-scale cosmological averages from an inhomogeneous space-time \cite{rev}. We will then proceed to show that the average cosmological properties that result from a general, inhomogeneous cosmological space-time cannot always be well represented by high-symmetry cosmological solutions of Einstein's equations. It is this mismatch that is at the heart of our proposed solution to the Hubble tension, and on which we will elaborate in what follows.

\subsection{Buchert's averaging formalism} \label{sec:buchert}

The Buchert averaging formalism is based on the 1+3-covariant decomposition of all quantities using the time-like 4-velocity of the irrotational flow of matter, $u^{\mu}$. Taking a covariant derivative of this 4-vector allows one to write:
\begin{equation} \label{du}
\nabla_{\nu} u_{\mu} = -u_{\nu} \dot{u}_{\mu} + \frac{1}{3} \vartheta h_{\mu \nu} + \varsigma_{\mu \nu} \, ,
\end{equation}
where $h_{\mu \nu} \equiv {\rm g}_{\mu \nu} + u_{\mu} u_{\nu}$ is the projection tensor onto the spaces orthogonal to $u^{\mu}$, and where a dot denotes the time derivative $u^{\mu} \nabla_{\mu}$. The expansion scalar is defined by $\vartheta \equiv h^{\mu \nu} \nabla_{\mu } u_{\nu}$, and the symmetric shear tensor is defined by $\varsigma_{\mu \nu} \equiv ( h_{( \mu}^{\phantom{( \mu} \rho} h_{\nu )}^{\phantom{\nu )} \tau} - \frac{1}{3} h_{\mu \nu} h^{\rho \tau} ) \nabla_{\rho} u_{\tau}$.

The spatial average of any scalar, $\mathcal{S}$, on the surfaces orthogonal to the fluid flow $u^{\mu}$ are then defined over a spatial domain $\mathcal{D}$ by
\begin{equation} \label{average}
\langle \mathcal{S} \rangle \equiv \frac{\int_\mathcal{D} \sqrt{h}\,  \mathcal{S}\, d^3 x}{\int_\mathcal{D}\sqrt{h}\, d^3 x} \, ,
\end{equation}
where $h$ is the determinant of $h_{\mu \nu}$. From this average, it immediately follows that the time derivative and averaging operators obey
$\langle \mathcal{S} \rangle \hspace{-0.15cm}\dot{\phantom{\mathcal{S}}} - \langle \dot{\mathcal{S}} \rangle = \langle \mathcal{S} \vartheta\rangle - \langle \mathcal{S} \rangle \langle \vartheta \rangle$.
Application of these operations to the Ricci identities $2 \nabla_{[\mu} \nabla_{\nu]} u^{\rho} = R_{\mu \nu \phantom{\rho} \tau}^{\phantom{\mu \nu} \rho} u^{\tau}$ and the Gauss embedding equation can be used to write
\begin{equation} \label{f1}
\frac{\dot{a}_\mathcal{D}^2}{a_\mathcal{D}^2} = \frac{8 \pi G}{3} \langle \mu \rangle+ \frac{\Lambda}{3} - \frac{\langle \mathcal{R} \rangle}{6}  - \frac{\mathcal{Q}}{6} \qquad {\rm and} \qquad \frac{\ddot{a}_\mathcal{D}}{a_\mathcal{D}} = -\frac{4 \pi G}{3} \langle \mu \rangle+ \frac{\Lambda}{3} + \frac{\mathcal{Q}}{3} \, ,
\end{equation}
where the matter content of the space-time has been taken to be dust with mass density $\mu$, where the scale factor $a_\mathcal{D}$ is defined implicitly by $\langle \vartheta \rangle= 3 \dot{a}_\mathcal{D}/a_\mathcal{D}$, and where $\mathcal{R}$ is the Ricci curvature scalar of the spaces orthogonal to $u^{\mu}$ (for which $h_{\mu \nu}$ is the induced metric). The average of the contracted second Bianchi identity gives $\langle \mu \rangle \propto {a}_\mathcal{D}^{-3}$.

The two equations in (\ref{f1}), and the average mass conservation equation, look remarkably similar in form to the Friedmann equations that govern the evolution of FRW models, but with the spatial curvature parameter replaced by $k \rightarrow a_\mathcal{D}^2 \langle \mathcal{R} \rangle/6$, and with additional terms containing the scalar
\begin{equation} \label{q}
\mathcal{Q} \equiv \frac{2}{3} \left( \langle \vartheta^2 \rangle - \langle \vartheta \rangle^2 \right) - 2 \langle \varsigma^2 \rangle \, ,
\end{equation}
where $\varsigma^2 \equiv \frac{1}{2} \varsigma_{\mu \nu} \varsigma^{\mu \nu}$. A necessary integrability condition for the equations above is
\begin{equation} \label{int}
\dot{\mathcal{Q}} + 6 \frac{\dot{a}_\mathcal{D}}{a_\mathcal{D}} \mathcal{Q} + \langle \mathcal{R} \rangle \hspace{-0.15cm}\dot{\phantom{\mathcal{S}}} + 2 \frac{\dot{a}_\mathcal{D}}{a_\mathcal{D}} \langle \mathcal{R} \rangle =0 \, .
\end{equation}
This immediately shows that in the special case in which $\mathcal{Q} \propto a_\mathcal{D}^{-6}$ the averaged equations (\ref{f1}) take the form of Friedmann's equations for a universe containing dust and a stiff fluid, or just dust if $\mathcal{Q}=0$. In these special cases $\langle \mathcal{R} \rangle \propto a_\mathcal{D}^{-2}$, and the averaged equations have a form that is identical to the Friedmann equations, meaning that at late times the large-scale averages from Buchert's formalism must behave in exactly the same way as the local equations of homogeneous and isotropic dust-dominated cosmologies. For all other $\mathcal{Q}$ we have $\langle \mathcal{R} \rangle$ not proportional to $a_\mathcal{D}^{-2}$, which we will refer to as `the general case'.

The averaged cosmological quantities discussed above are not direct observables. They are, however, direct representations of the kinematical and curvature properties of extended regions of space. For example, the average expansion scalar $\langle \vartheta \rangle$ gives the rate of volume expansion of the spatial domain $\mathcal{D}$, while the scale factor $a_{\mathcal{D}}$ directly represents its spatial scale \cite{buchert1, buchert2}. Though not directly observable, such quantities are of clear interest for understanding the large-scale properties of a cosmological space-time. To make inferences on cosmological quantities from astronomical observables, however, one needs to go further and relate these averages to the observables that are recorded by our telescopes (i.e. redshifts, luminosity distances, etc). This will be the subject of Section \ref{sec:optics}.

\subsection{Cosmological models with symmetries} \label{cossym}

Let us now consider the space-times that are solutions of Einstein's equations, and that admit a foliation with leaves that have a spatial domain $\mathcal{D}$ that has at every point:
\begin{itemize}
\item[(i)] an isotropic part of their expansion rate, $\theta$, equal to $\langle \vartheta \rangle$.
\item[(ii)] a Ricci curvature scalar, ${}^3\hspace{-0.05cm}R$, equal to $\langle \mathcal{R} \rangle$. 
\end{itemize}
In Appendices A and B we show that in the general case there exist no explicitly homogeneous and isotropic space-times that can obey both conditions (i) and (ii), as these geometries simply do not allow sufficient freedom to simultaneously reproduce the averages of the rate of expansion and curvature of space that can be produced within general inhomogeneous cosmologies. Furthermore, there can exist no regular hypersurface-orthogonal and spherically-symmetric barotropic perfect-fluid solutions of Einstein's equations that can maintain the averaged values of these quantities at all points in space, as well as no untilted homogeneous perfect-fluid cosmologies of Bianchi types I or V (the simplest anisotropic cosmologies). These results severely limit the high-symmetry solutions of Einstein's equations that one could use to fit the large-scale properties of the Universe in the general case, and strongly suggests that we should {\it not} be using such space-times as models of a general inhomogeneous universe (an idea previously discussed in Refs.~\cite{ras1, ras2}). This is counter to assumption (2) from the introduction, as well as the vast majority of current approaches to cosmological modelling.

We are drawn to the conclusion that we should think of large-scale averages such as $\langle \vartheta \rangle$ and $\langle \mathcal{R} \rangle$ as simply being statistical properties of space at some moment of time, rather than as corresponding directly to the properties of any cosmological solution of Einstein's equations with explicit symmetries, as in general the latter simply do not exist. If we choose, we can instead think of the results of spatial averaging as constituting a new type of cosmological model that consists of a set of intrinsically symmetric 3-spaces with properties spatial curvature $\langle \mathcal{R} \rangle$ and expansion $\dot{a}_{\mathcal{D}}/a_{\mathcal{D}}$. This is in keeping with assumption (1) from the introduction, and demonstrates that this is the weaker of the two assumptions. It is, however, a considerable shift in perspective, as the resulting cosmological model should no longer be thought of as having the properties of a space-time in its own right. This is not to say that the real inhomogeneous Universe is not a proper space-time, satisfying Einstein's equations, just that the family of intrinsically symmetric 3-spaces that form our new cosmological model should not be expected to have this property. As we will see in the next sections, this change in perspective could have significant consequences for interpreting the Hubble tension.

\section{Optical properties and fitting models} \label{sec:optics}

In this section we will consider the optical properties of inhomogeneous space-times, and the relationships between the averages of observable quantities, such as distance measures and redshifts, and the averaged cosmological quantities from Section \ref{sec:buchert}. This will all be done in statistically homogeneous and isotropic space-times, and will closely follow the work of R$\ddot{\rm a}$s$\ddot{\rm a}$nen \cite{ras1, ras2}. We will then consider the cosmological parameters that an observer would infer if they interpeted these observations within a spatially flat FRW fitting model (to be defined below).

\subsection{Optical properties}

The optical properties of statistically homogeneous and istropic cosmological models have been considered in a very general sense by R$\ddot{\rm a}$s$\ddot{\rm a}$nen in Refs. \cite{ras1, ras2}. This work starts from the Sachs optical equations \cite{sachs}, which are a set of evolution equations for the expansion $\tilde{\theta}$, shear $\tilde{\sigma}_{\mu \nu}$ and vorticity $\tilde{\omega}_{\mu \nu}$ of congruences of null geodesics, derived under the geometric optics approximation, but otherwise not restricted to any particular space-time. From these quantities, the angular diameter distance and redshift of a source are given by
\begin{equation}
D_A \propto {\rm exp} \left( \frac{1}{2} \int {\rm d} \lambda \, \tilde{\theta} \right)  \qquad {\rm and} \qquad 1+z = {\rm exp} \left( \int {\rm d} \lambda \, E \, \left[\frac{1}{3} \vartheta + \varsigma_{\mu \nu} e^{\mu} e^{\nu} \right] \right) \, ,
\end{equation}
where $\lambda$ is an affine parameter along the given bundle of light rays, $E= - u_{\mu} k^{\mu}$ is the energy of the photon, $\vartheta$ and $\varsigma_{\mu\nu}$ are the expansion and shear of matter fields, and $e^{\mu}$ is a space-like unit vector orthogonal to the matter flow $u^{\mu}$ and pointing in the direction of the ray of light, and where Etherington's theorem is valid along every ray of light \cite{eth}. After a careful treatment, R$\ddot{\rm a}$s$\ddot{\rm a}$nen argues that in dust-filled space-times we should expect that observations made over cosmological scales should recover \cite{ras1}
\begin{equation}
\langle D_A \rangle \propto {\rm exp} \left( \frac{1}{2} \int {\rm d} \lambda \, \langle \tilde{\theta} \rangle \right)
\qquad {\rm and} \qquad
1+ \langle z \rangle = \frac{1}{a_{\mathcal{D}}} \, ,
\end{equation}
where $a_{\mathcal{D}}$ is the scale factor that appears in the averaged Friedmann equations (\ref{f1}). Differentiation of these results, and application of the Sachs equations, then gives \cite{ras1}
\begin{equation} \label{raseq}
H \frac{\partial}{\partial \langle z \rangle} \left[ (1+ \langle z \rangle)^2 H \frac{\partial}{\partial \langle z \rangle} \langle D_A \rangle \right] = - \left[ 4\pi G \langle \mu \rangle +\frac{\langle \tilde{\sigma}^2\rangle}{\langle E \rangle^2} \right] \, \langle D_A \rangle \, ,
\end{equation}
where $H= \dot{a}_{\mathcal{D}}/a_{\mathcal{D}}$ is the large-scale expansion of space, and $\tilde{\sigma}^2 = \frac{1}{2} \tilde{\sigma}_{\mu \nu} \tilde{\sigma}^{\mu \nu}$ is the magnitude of the null shear. For investigations of this result, including the effects of shear in the matter flow, see e.g. \cite{syksy2, kok2, kok3, kok4}.

In most applications in the real Universe the effects of null shear are expected to be small \cite{shear1}, so in the present study will neglect the contribution of $\langle \tilde{\sigma}^2\rangle$. In this case we can manipulate Eq. (\ref{raseq}) into the following form:
\begin{equation} \label{d}
\ddot{D} + \frac{\dot{a}_{\mathcal{D}}}{a_{\mathcal{D}}} \dot{D} + \left( \frac{\langle \mathcal{R} \rangle}{6}+ \frac{\mathcal{Q}}{2} \right) D =0 \, ,
\end{equation}
where over-dots again denote differentiation with respect to time, $u^{\mu} \nabla_{\mu}$, and where we have defined $D\equiv (1+\langle z \rangle) \langle D_A \rangle$. This equation is a second-order ordinary differential equation, which can be solved using the following initial conditions: $D(t_0) = 0$ and $\dot{D}(t_0) = -1$, where $t_0$ is the present time. It can be seen that Eq. (\ref{d}) has two forcing terms that drive the evolution of $D$; one from the averaged spatial curvature $\langle \mathcal{R} \rangle$, and one from the scalar $\mathcal{Q}$. The first of these is present in standard FRW cosmologies, and is responsible for the strong constraints that CMB measurements impose upon spatial curvature. The second is not present in FRW cosmology, but can be seen to enter into Eq. (\ref{d}) in a similar way to the spatial curvature. Eq. (\ref{d}) explicitly demonstrates the link between averaged observables and the quantities $\langle \mathcal{R} \rangle$ and $\mathcal{Q}$, which arise from the spatial averaging process in Section \ref{sec:buchert}.

In what follows we will solve the distance equation (\ref{d}), together with the Friedmann equations (\ref{f1}) and the integrability condition (\ref{int}), after specifying some example forms for $\mathcal{Q}$. The results of this will give us the expansion history of our new cosmological models, in terms of the scale factor and Hubble expansion at each instant of time, along with the distance measure $D$. Our aim will not be to try and extract $\mathcal{Q}(t)$ from any particular model of structure formation, as this is an extremely complicated subject that has proven to be controversial in the literature on this subject, with different approaches yielding different results (see e.g. \cite{br1, br2, br3, br4}). Instead, we will consider some example functional forms of $\mathcal{Q}(t)$ that we believe yield interesting results in the context of the Hubble tension.

\subsection{Fitting models}

With $\mathcal{Q}(t)$ specified, and average observables calculated, we wish to consider the parameters that an observer would infer if they used spatially flat FRW as a fitting model for the Hubble diagrams they observe in their inhomogeneous universe. The idea here is that observer would be trying to extract all the information they can from the observations they record, under the assumption the spatially flat FRW model is a good description of the universe in which they live. The ``fitting model'' in this process is the FRW model, which we are distinguishing from the ``cosmological models'' described in Section \ref{sec:modelling}, as in general FRW models are not capable of capturing the full range of behaviours that can be displayed in a general inhomogeneous cosmology (see Section \ref{cossym}). That is, the ``fitting model'' should be considered to be the model that the observer fits to their observational data, but that does not necessarily relate directly to the average properties of the underlying space-time (see Ref. \cite{fitting} for further discussion).

Our approach to this problem will be to assume that our observers have access to perfectly precise idealised observations of $D(z)$, and that they fit a spatially-flat FRW model to this data. We will then extract the values of the energy density $\rho$ and the isotropic pressure $p$ that this observer would infer from their fitting model, as well as the Hubble rate as a function of redshift $H(z)$. This last quantity will be particularly important for the Hubble tension, and in the next section we will show that in the general case even an idealised observer with access to perfect observations may not infer the correct Hubble rate; in general, they would infer different $H$ from different observables, and hence be left with a Hubble tension.

To extract the Hubble rate that an observer would infer from an FRW fitting model of the type described above, we can note that from Eq. (\ref{d}) the spatially-flat FRW limit tells us that $a_{\mathcal{D}} \dot{D}$ is a constant in time. This then gives the Hubble rate of the fitted FRW model as
\begin{equation}
H_{\rm FRW} = \frac{\rm constant}{{\rm d} D/{\rm d} z} \, ,
\end{equation}
where the $D$ in this equation would be the one determined from the real inhomogeneous universe, and where the constant should be set by initial conditions. With the expansion rate of the FRW fitting model determined, we can then extract the energy density and pressure in the fitting model using
\begin{equation}
\rho_{\rm FRW} = \frac{3 H_{\rm FRW}^2}{8 \pi G} \qquad {\rm and} \qquad p_{\rm FRW} = \frac{1}{3} (1+z) \frac{{\rm d}\rho_{\rm FRW}}{{\rm d} z} - \rho_{\rm FRW} \, ,
\end{equation}
which immediately gives the equation of state of the hypothetical fluid in the fitting model as $w_{\rm FRW} = {p_{\rm FRW}}/{\rho_{\rm FRW}}$. In general, none of these FRW quantities should be expected to accurately represent the averages in the inhomogeneous universe, and in particular we should expect $H_{\rm FRW} \neq H$. It is the difference between these two quantities that will be the focus of the next section, and which we interpret as being responsible for the Hubble tension.

\section{Hubble tension and the radial BAO} \label{sec:results}

In this section we will use some simple example forms for the function $\mathcal{Q}$, and show that an observer in a universe with such a function, who interpreted their astronomical data using an FRW model, would infer an incorrect difference in Hubble rate between last scattering and today. As a corollary of this hypothesis, we make a prediction for observations of the radial component of the Baryon Acoustic Oscillations (BAO), and suggest that a signature of this prediction appears to exist in the recent DESI first-year data release \cite{desi1, desi2}.

\subsection{Hubble tension}

\begin{figure}
    \centering
    \begin{subfigure}[b]{\textwidth}
        \centering
        \includegraphics[width=15cm]{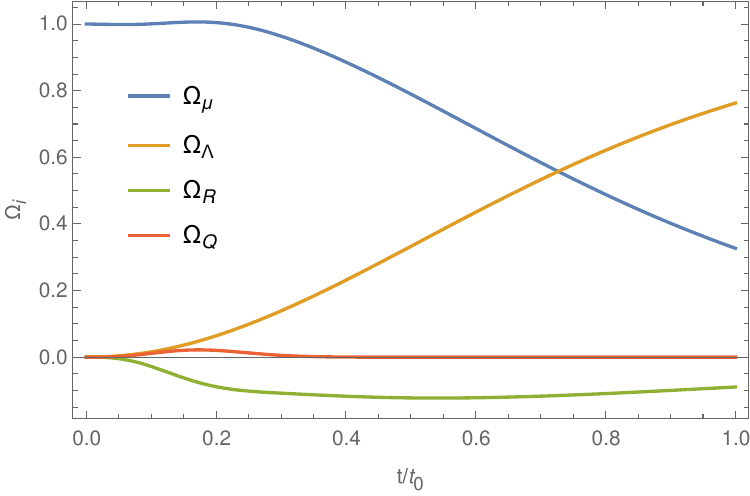}
        \newline
        \vspace{-0.75cm}
        \flushleft{Model (a): a cosmology with $\mathcal{Q}<0$ and $\Lambda >0$, in which a small negative $\mathcal{Q}$ at early times sources positive $\langle \mathcal{R} \rangle$.}
    \end{subfigure} 
    \\[10pt]
    \begin{subfigure}[b]{\textwidth}
        \centering
        \includegraphics[width=15cm]{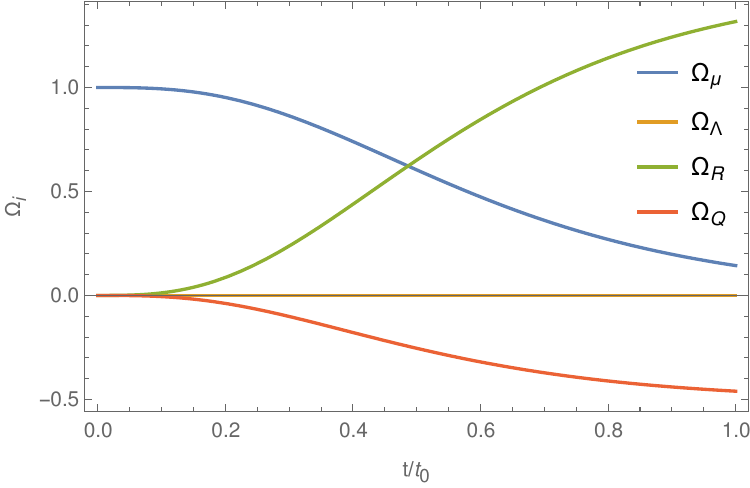}
        \newline
        \vspace{-0.75cm}
        \flushleft{Model (b): a cosmology with $\mathcal{Q}>0$ and $\Lambda =0$, in which a large positive $\mathcal{Q}$ sources negative $\langle \mathcal{R} \rangle$ at late times, which acts like dark energy.}
    \end{subfigure}
    \caption{The density parameters $\Omega_i$ for the example cosmological models (a) and (b).}\label{negQandDE}
\end{figure}

The functional forms of the magnitude $Q$, for the example cases we will consider, are shown as red lines in Fig. \ref{negQandDE}, where we display the density parameters $\Omega_i$ defined by
\begin{equation}
\Omega_{\mu} = \frac{8 \pi G \langle \mu \rangle}{3 H^2} \, , \qquad
\Omega_{\Lambda} = \frac{\Lambda}{3H^2} \, , \qquad
\Omega_{\mathcal{R}} = - \frac{\langle \mathcal{R} \rangle}{6 H^2} \, , \qquad
\Omega_{\mathcal{Q}} = - \frac{\mathcal{Q}}{6 H^2} \, .
\end{equation}
Model (a), shown in the upper plot, has $\Omega_{\mathcal{Q}}$ with a positive value (corresponding to a negative $\mathcal{Q}$), and a maximum amplitude at $t\simeq 0.15 t_0$. This model corresponds to a universe that contains a positive cosmological constant $\Lambda$, and develops some positive spatial curvature, which subsequently decays. In this model the value of $\Lambda$ is chosen such that $\Omega_{\Lambda}/\Omega_{\mu} = 0.7/0.3$ at $t=t_0$. Model (b), shown in the lower plot, has $\Omega_{\mathcal{Q}}$ with a negative value (corresponding to a positive $\mathcal{Q}$), and grows in amplitude all the way up to $t=t_0$. This model has $\Omega_{\Lambda}=0$, as in this case the growing negative spatial curvature is sufficient to cause apparent acceleration without a cosmological constant.

These two models are not chosen because we necessarily expect them to correspond to the physical situation in the real Universe, but instead because they display properties that are of interest for understanding how a Hubble tension could arise in the scenario we are describing. They are intentionally extreme examples, chosen only to make a conceptual point. The behaviour exhibited in model (a) can be understood by recognising that the integrability condition (\ref{int}) means that a non-zero $\mathcal{Q}$ sources a spatial curvature $\langle \mathcal{R} \rangle$ with the opposite sign. When the value of $\mathcal{Q}$ decays back to zero, as $t \rightarrow t_0$, the spatial curvature remains, and slowly decays as $\langle \mathcal{R} \rangle \propto a_{\mathcal{D}}^{-2}$. The effect of the positive spatial curvature, in this case, is to shift the true value of the Hubble rate away from that of the FRW fitting model, which is why we have included it. In model (b) the situation is somewhat different, as the value of $\mathcal{Q}$ is large and positive. In this case the spatial curvature $\langle \mathcal{R} \rangle$ is still being sourced, but it is not the dominant contribution to the difference in Hubble rates between the true cosmology and the FRW fitting model. Instead, in model (b) it is $\mathcal{Q}$ itself that is the primary contributor to the Hubble tension, through its presence in Eq. (\ref{f1}). Our two models therefore illustrate two different ways of creating a Hubble tension.

\begin{figure}
\center
\includegraphics[width=15cm]{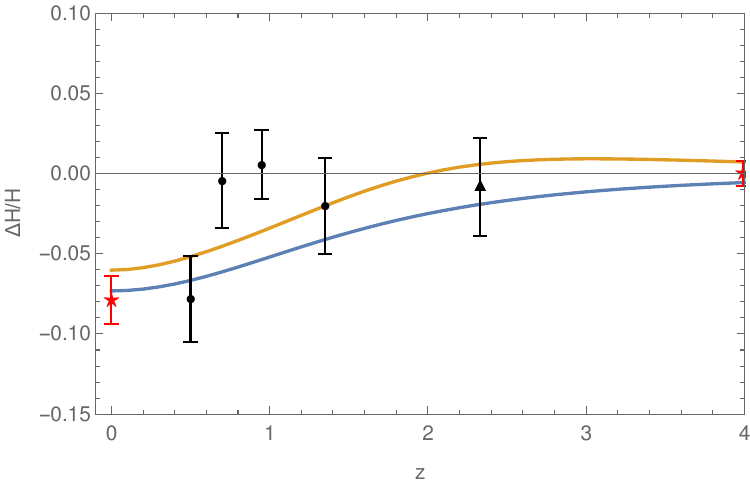}
\caption{The Hubble tension, $\Delta H/H$, inferred by an observer who precisely models the angular diameter distance using an FRW fitting model. The orange curve is model (a), and the blue curve is model (b). The red stars represent the Hubble tension between Planck \cite{cmb} (displayed on the RHS of the diagram) and SHOES \cite{shoes} (displayed at $z=0$). The black circles correspond to the DESI first-year data from galaxies and quasars \cite{desi1}, and the black triangle to the data from the Lyman-$\alpha$ forest \cite{desi2}, as discussed in Section \ref{apsec}.}  \label{dH}
\end{figure}

Within our example models we can now solve Eq. (\ref{d}) to find the angular diameter distance as a function of redshift, $D_A(z)$, and use Etherington's theorem to find the luminosity distance, $D_L(z)$. An FRW model can then be fitted to these distance-redshift relations, though we note that different observables may require fitting in different ways. For example, supernovae observations probe the luminosity distances at relatively low redshifts, and can therefore be used to measure the local Hubble rate quite directly through the leading-order approximation $H_0 \simeq z/D_L$, which from Section \ref{sec:optics} can be seen to be equal to $\dot{a}_{\mathcal{D}}/a_{\mathcal{D}}$ (assuming we live in a typical region of the Universe). On the other hand, for the CMB we need to require that the angular diameter distance to the last-scattering surface at $z=1100$ takes the same value in the fitting model as it does in the actual cosmology, in order for the angular scale observed in the CMB fluctuations to be correctly reproduced. With the corresponding FRW fitting model in hand, we can then determine the precise difference between the Hubble rate at last scattering and that at any $z \lesssim 1100$ in both the erroneous FRW model and the true underlying inhomogeneous cosmology. To quantify the difference in Hubble rate between last scattering at $t=t_{\rm LS}$, and some time $t=t(z)$, we define
\begin{equation}
\delta H (z) \equiv \frac{H(t(z))}{H (t_{\rm LS})} \qquad {\rm and} \qquad \delta H_{\rm FRW} (z) \equiv \frac{H_{\rm FRW}(t(z))}{H_{\rm FRW} (t_{\rm LS})}\,,
\end{equation}
and then
\begin{equation}
\frac{\Delta H}{H} \equiv \frac{\delta H_{\rm FRW}- \delta H}{\delta H} \, ,
\end{equation}
as the error inferred by using a spatially flat FRW universe to fit the CMB, where $H=\dot{a}_{\mathcal{D}}/a_{\mathcal{D}}$. This quantity is displayed in Fig. \ref{dH} for our two models, with the consequential point with respect to the Hubble tension being given by the value of $\Delta H/H$ at $z=0$. In both cases, the difference in Hubble rate between last scattering and today is over-estimated in the FRW model, with $\Delta H <0$ at low $z$. We suggest that it may be an inference error of this kind that has led to the Hubble tension in the real Universe.

The two models discussed in this section provide two different mechanisms by which a genuine Hubble tension could be produced. However, they also predict that the true value of the Hubble rate, and the Hubble rate inferred from using an FRW fitting model, should be different at redshifts $z > 0$ (the idea that the difference in Hubble rates should exist only at the present time is clearly implausible). The precise value of the Hubble tension at $z \geq 0$ depends on the form of $\mathcal{Q}$, as can be seen by the two curves corresponding to our two different models taking different shapes in Fig. \ref{dH}. This provides us an opportunity to test our hypothesis, by considering observables that are directly sensitive to the Hubble rate at the different redshifts (though one needs to be careful to ensure it is the true Hubble rate that the observable depends upon, and not the inferred one from the FRW model). One such observable is the radial component of the BAO observations, which is the subject of the next section.

\subsection{The radial BAO} \label{apsec}

The BAO feature can be measured in two different ways; either using the angular diameter distance in order to determine the physical length scale corresponding to an angle on the sky of the observer, or through the difference in redshift that is used to infer a physical radial distance. The latter depends on the Hubble rate in order to get a representative physical length scale. If the Hubble rate in the real Universe is different to the Hubble rate in the fitting model, then the radial BAO scale will consequently appear different from expectations based on FRW cosmology. In order to discuss the radial BAO, we can start by defining
\begin{equation}
\alpha_{\parallel}(z) \equiv \frac{H^{\rm fid} r_{\rm drag}^{\rm fid}}{H \, r_{\rm drag}} \simeq \frac{H^{\rm fid}}{H}
\, ,
\end{equation}
which is the isotropic term in the general expression for this quantity \cite{umeh}. Here, $r_{\rm drag}$ is the comoving sound horizon, evaluated at the redshift at which baryon-drag optical depth equals unity, and where objects labelled with a superscript ``fid'' are to be evaluated in the fiducial FRW model used to change redshifts into distances in the survey. In the last equality we have assumed that $r_{\rm drag}\simeq r_{\rm drag}^{\rm fid}$.

Within our class of models we can therefore interpret the constraints on $\alpha_{\parallel}$ in terms of the induced Hubble tension at a given redshift, with
\begin{equation}
\frac{\Delta H}{H} (z) \simeq \alpha_{\parallel} (z) -1 \, .
\end{equation}
Using the information from the DESI first-year galaxy and quasar results \cite{desi1}, and the Lyman-$\alpha$ forest \cite{desi2}, we can therefore add some further data points to Fig. \ref{dH}. We do this using black dots and triangles, respectively, for the two sets of observables. There are some clear discrepancies between the observations and the fiducial model, which in our present framework is an indication of a Hubble tension at that redshift. From our Fig. \ref{dH}, the level of discrepancy appears very similar to the datum at $z=0$ that corresponds to the Hubble tension (though with a larger error bar). This is exactly as one would expect, if both discrepancies have the same origin. This can be contrasted with the SDSS data \cite{sdss2}.

\section{Discussion} \label{sec:disc}

Taking the Hubble tension seriously, one is presented with a crisis in modern cosmology, in which the favoured cosmological model results in discordant constraints on the same parameter at high statistical significance. We suggest that this discordance may be due to the cosmological model being overly simplistic, and suggest a radical alternative in which statistical homogeneity and isotropy of space are {\it not} taken to imply that our cosmological model should be close to a single space-time which at all points exhibits explicit symmetry. We have shown that this view is necessary if one wishes to consider a cosmological model in which spatial curvature develops, as has been suggested by a variety of approaches to extracting averages from inhomogeneous cosmological space-times \cite{buchert1, buchert2, zal, sang1, sang2}. In such cases, the large-scale expansion of space, and distance measures within that space, are no longer tied by the same relationships, and a Hubble tension between inferences made using different sets of observables can naturally arise.

We have demonstrated our hypothesis within Buchert's averaging framework \cite{buchert1, buchert2}, using the distance measures derived by R$\ddot{\rm a}$s$\ddot{\rm a}$nen for application in statistically homogeneous and isotropic cosmologies \cite{ras1, ras2}. While the cosmological behaviours required to relax the Hubble tension are very considerable departures from the standard approach to cosmology, we find multiple ways in which this relaxation can be achieved. As a corollary of our hypothesis, we also suggest that inferences on radial BAO parameter $\alpha_{\parallel}$ may be linked to the Hubble tension, and suggest that this link may in fact have already been detected in the DESI first-year data release \cite{desi1, desi2}, which has so far been interpreted as corresponding to dynamical dark energy \cite{desi3}. While the Hubble tension cannot be used to reconstruct $\alpha_{\parallel}$ at all values of $z$, it does provide an anchor for the expected value at $z=0$. Conversely, if BAO observations become good enough, it may be possible to reconstruct $\Delta H/H$ as a function of $z$ explicitly, and use this to infer compatible values of the Hubble tension that astronomers should be expected to measure using distance ladders. Such an approach may allow for the possibility of reconstructing $\mathcal{Q}(z)$ from observations in the real Universe, rather than trying to extract it from simulations only.

Our work differs from the vast majority of other approaches to solving the Hubble tension problem in that it does not adjust how physics works in an FRW cosmology, but rather changes the nature of what we consider a cosmological model to be by removing the association with a space-time with explicit symmetries. In this regard it is subtly different from previous work on the Hubble tension in averaged cosmology performed in Ref. \cite{ten1}, in that it does not require the introduction of a `template metric' \cite{temp1, temp2}. It can also be compared with the work in Ref. \cite{ten2}, which was performed in the context of comparing expansion rates at different cosmological times in a simulation of a `silent universe', and with the work in Ref. \cite{ten3} which discusses the idea of the Hubble tension arising as a manifestation of the fitting problem. Finally, we note that our scenario has the side-effect of reducing the age of the Universe, which could have consequences for the $S_8$ tension, as well as constraints from the ages of astrophysical objects \cite{inter}. In future works, we hope to perform a detailed fitting to a suite of cosmological observables, as in e.g. \cite{syksy3, syksy4, kok1}, in order to further investigate the plausibility of this idea as a solution to the Hubble tension problem.

{\flushleft
{\bf Acknowledgements:} We gratefully acknowledge helpful conversations, comments and suggestions from Jenny Wagner, Roy Maartens, Chris Clarkson, Karim Malik, Thomas Buchert, Asta Heinesen, Syksy R$\ddot{\rm a}$s$\ddot{\rm a}$nen, Obinna Umeh and Sofia Koksbang.}

\section*{Appendix A: Explicitly homogeneous geometries}

Let us try and find an explicitly homogeneous space-time geometry, obeying conditions (i) and (ii) from Section \ref{cossym}. To do this, we define a time-like congruence whose tangent vectors have components $n^{\mu}$. If this congruence is orthogonal to homogeneous 3-spaces then it must be irrotational, and can therefore be covariantly decomposed as
\begin{equation}
\nabla_{\nu} n_{\mu} = n_{\nu} \, \dot{n}_{\mu} + \frac{1}{3} \theta f_{\mu \nu} + \sigma_{\mu \nu} \, ,
\end{equation}
where $f_{\mu \nu} = g_{\mu \nu}+n_{\mu} n_{\nu}$ is the projection tensor associated with $n^{\mu}$, and where an over-dot now denotes the time derivative $n^{\mu} \nabla_{\mu}$. We used the symbols $\{ g_{\mu \nu}, \theta , \sigma_{\mu \nu} \}$ for the idealized homogeneous geometry here, to distinguish them from those of the real inhomogeneous universe $\{{\rm g}_{\mu \nu} , \vartheta , \varsigma_{\mu \nu} \}$ used above. The expansion scalar and shear of the homogeneous geometry are, however, defined analogously.

The Gauss embedding equation and the Raychaudhuri equation, together with Einstein's equations, allow us to then write the following for the homogeneous geometry:
\begin{eqnarray}
&&{}^3\hspace{-0.05cm}R= 16 \pi \, G \, \rho - \frac{2}{3} \theta^2 + 2 \sigma^2 + 2\Lambda  \label{hf1}\\
&&\dot{\theta} +\frac{1}{3} \theta^2 = - 4 \pi \, G\, (\rho +3 p) - 2 \sigma^2 + \Lambda + \tilde{\nabla}_{\mu} \dot{n}^{\mu} +\dot{n}_{\mu} \dot{n}^{\mu} \, , \label{hf2}
\end{eqnarray}
where $\sigma^2 = \frac{1}{2} \sigma_{\mu \nu} \sigma^{\mu \nu}$, and where $\rho = T_{\mu \nu} n^{\mu} n^{\nu}$ and $p=\frac{1}{3} T_{\mu \nu} f^{\mu \nu}$ are the energy density and isotropic pressure of the stress-energy tensor $T_{\mu \nu}$ in the homogeneous model. The spatially projected derivative $\tilde{\nabla}_{\mu}$ is defined such that e.g. $\tilde{\nabla}_{\mu} \dot{n}_{\nu} = f_{\mu}^{\phantom{\mu} \alpha} f_{\nu}^{\phantom{\nu} \beta} {\nabla}_{\alpha} \dot{n}_{\beta}$.

By comparing the homogeneous equation (\ref{hf1}) to the first of the averaged equations in (\ref{f1}), and applying the two conditions ${}^3\hspace{-0.05cm}R=\langle {}^3\mathcal{R} \rangle$ and $\theta = \langle \vartheta \rangle$, we can immediately read off that we require
$
\mathcal{Q} = 16 \pi \, G \, \left(\langle \mu \rangle - \rho\right) - 2 \sigma^2
$.
Subsequently, comparing Eq. (\ref{hf2}) and the second equation in (\ref{f1}), and using the result above, implies that we must have
\begin{equation} \label{hp}
8 \pi \, G \, p = - 8 \pi \, G \, (\rho -\langle \mu \rangle) + \frac{2}{3} \tilde{\nabla}_{\mu} \dot{n}^{\mu} + \frac{2}{3} \dot{n}_{\mu} \dot{n}^{\mu} \, ,
\end{equation}
where we have assumed that the time-derivative operators in the averaged and idealized homogeneous geometries are identical. We can then use the shear evolution equation derived from the Ricci identities and Gauss embedding equation to write
\begin{equation} \label{s2}
(\sigma^2)\dot{} + 2 \theta \sigma^2 = \sigma_{\mu \nu} \pi^{\mu \nu} -  {}^3\hspace{-0.05cm}R^{* \mu \nu}\sigma_{\mu \nu} + \sigma^{\mu \nu} \tilde{\nabla}_{\mu} \dot{n}_{\nu} + \sigma_{\mu \nu} \dot{n}^{\mu} \dot{n}^{\nu} \, ,
\end{equation}
where $\pi_{\mu \nu} = f_{(\mu}^{\phantom{(\mu} \alpha}  f_{\nu)}^{\phantom{\nu)} \beta} T_{\mu \nu} - \frac{1}{3} f_{\mu \nu} f^{\alpha \beta} T_{\alpha \beta}$ is the anisotropic stress tensor, and ${}^3\hspace{-0.05cm}R^*_{\mu \nu}$ is the trace-free part of the Ricci curvature tensor of the 3-spaces orthogonal to $n^{\mu}$. Now, if we were to have $\pi_{\mu \nu}={}^3\hspace{-0.05cm}R^*_{\mu \nu}=0 = \dot{n}^{\mu}$ then the right-hand side of Eq. (\ref{s2}) would vanish, and we would immediately have $\sigma^2 \propto a^{-6}$, where the scale factor $a$ is defined implicitly by $\theta= 3 \dot{a}/a$. Under these same conditions, and taking $\langle \vartheta \rangle = \theta$, the energy conservation equations for $\langle \mu \rangle$ and $\rho$ imply
\begin{equation}
(\rho -\langle \mu \rangle)\dot{} + 2 \theta (\rho -\langle \mu \rangle) = \tilde{\nabla}_{\mu} q^{\mu} \, ,
\end{equation}
where $q^{\mu} = - T_{\alpha \beta} n^{\alpha} f^{\beta \mu}$ is the momentum density. This means that if $q^{\mu}=0$ then we also have $(\rho- \langle \mu \rangle) \propto a^{-6}$, and hence that $\mathcal{Q} \propto a^{-6}$. This means that any homogeneous geometry with $\pi_{\mu \nu}={}^3\hspace{-0.05cm}R^*_{\mu \nu}=0 = \dot{n}^{\mu} = q^{\mu}$ cannot represent the general case. As $\pi_{\mu \nu}={}^3\hspace{-0.05cm}R^*_{\mu \nu}=0 = \dot{n}^{\mu} = q^{\mu}$ in every homogeneous and isotropic geometry, this means that no homogeneous and isotropic geometries can be used to model the general case. Furthermore, as all homogeneous perfect fluids have $\dot{n}^{\mu}=0$, and as spatial curvature is isotropic in all Bianchi type I and V geometries, we can also note that there are no hypersurface-orthogonal perfect-fluid Bianchi type I or V geometries that can be used to model the general case either.

\section*{Appendix B: Spherically symmetric geometries}

Reducing the symmetry requirements, it is possible to consider spherically symmetric geometries. While these spaces will not satisfy the Copernican principle in the same way as the homogeneous spaces from Appendix A, they do allow for the construction of a space-time that is isotropic around one particular point; the centre of symmetry. They could, therefore, potentially act as cosmological geometries for observers positioned at that point.

Spherically symmetric cosmological models are well studied, and lend themselves to the methods used to study locally-rotationally-symmetric (LRS) solutions of Einstein's equations \cite{lrs}. In what follows we will follow the semi-tetrad approach used to systematically study LRS cosmologies in Ref. \cite{elst}. Here, the preferred space-like direction corresponding to the axis of rotational symmetry is given by $e^{\mu}$, defined such that $e_{\mu} n^{\mu} = 0$ and $e_{\mu} e^{\mu}=1$. The acceleration, shear and electric parts of the Weyl tensor can then be written in the semi-tetrad $\{n^{\mu}, e^{\mu}\}$ as
\begin{equation}
\dot{n}^\mu = \dot{n} \, e^{\mu} \, , \qquad
\sigma_{\mu \nu} = \frac{2}{\sqrt{3}} \, \sigma \, e_{\mu \nu} \qquad {\rm and} \qquad
E_{\mu \nu} = \frac{2}{\sqrt{3}} \, E \, e_{\mu \nu} \, ,
\end{equation}
where $E_{\mu \nu} = n^{\sigma} n^{\rho} C_{\mu \sigma \nu \rho}$ and $e_{\mu \nu} = \frac{1}{2} (3 e_{\mu} e_{\nu} - f_{\mu \nu})$, and where $C_{\mu \sigma \nu \rho}$ is the Weyl tensor. Spherically symmetric space-times are of LRS class II and, in addition to the quantities written above, require only the space-like expansion $a= f^{\mu}_{\phantom{\mu} \nu} \nabla_{\mu} e^{\nu}$ in order to be specified.

Now, for a perfect fluid with $q^{\mu}=0=\pi_{\mu\nu}$ the Ricci identities imply the constraint \cite{elst}
\begin{equation}
\sigma'+ \frac{3}{2} \, a\, \sigma = \frac{1}{\sqrt{3}} \,\theta' \, ,
\end{equation}
where primes denote differentiation in the preferred space-like direction, $e^{\mu} \nabla_{\mu}$. If we require of our spherically symmetric geometry that it has the same value of $\theta$ at each point in space, and that $\sigma=0$ at the centre of symmetry, then we can immediately see from the equation above that we must have $\sigma=0$ at all points. Solutions of this type have been considered in Refs. \cite{wyman, exact}.

In the present case, it can be seen that the shear-free restriction means that the back-reaction scalar must take the form $\mathcal{Q} =16 \pi \, G\,( \langle \mu  \rangle -\rho)$. From the averaged equations in (\ref{f1}), and the requirements that the spherically symmetric geometry takes the values of $\theta=\langle \vartheta \rangle$ and ${}^3\hspace{-0.05cm}R =\langle \mathcal{R} \rangle$ at all points in space, we are then left with the result that we must have a homogeneous energy density. For a perfect fluid with barotropic equation of state, $p=p(\rho)$, the Bianchi identities then imply for all $p \neq -\rho$ we must have \cite{elst}
\begin{equation}
\dot{n} = -\frac{\partial p }{\partial \rho} \frac{\rho'}{(\rho +p)} =0 \, .
\end{equation}
Using the results $\sigma = \dot{n}=0$, the evolution equation for the spatial curvature scalar becomes
\begin{equation}
({}^3\hspace{-0.05cm}R)\dot{} + \frac{2}{3} \, \theta \, {}^3\hspace{-0.05cm}R = 0 \, ,
\end{equation}
which using the integrability condition (\ref{int}) implies $\mathcal{Q} \propto a^{-6}$. This means that there are no regular hypersurface-orthogonal spherically-symmetric cosmological models with barotropic perfect fluids that can model the general case either. In fact, it can be seen from the Bianchi identities, which imply the constraint \cite{elst}
\begin{equation}
E'+\frac{3}{2} \, a \, E = \frac{4 \pi G}{\sqrt{3}} \, \rho' \, ,
\end{equation}
that the only regular solutions with $E=0$ at the centre of symmetry, and $\rho'=0$ at all points, must have $E=0$ everywhere. This means that all such geometries are in fact FRW models, and therefore that the only spherically symmetric solutions that could possibly model the general case are those that contain fluids that are either (i) imperfect, (ii) not hypersurface orthogonal, and/or (iii) not barotropic. 

\section*{Appendix C: Summary of notation}

Our presentation requires the use of similarly named quantities applied in different situations, so we present here a table summarizing the notation we have used in different contexts, to help guide the reader. In particular, we have defined kinematic and curvature quantities in referring to (i) the inhomogeneous geometry of the Universe, (ii) the geometry of idealized cosmological models, and (iii) the optical properties of the Universe as follows:

\begin{center}
\begin{tabular}{| l || c | c | c |} 
 \hline
\phantom{\LARGE{A}} & expansion & shear & scalar curvature \\ [0.5ex] 
 \hline\hline
 (i) underlying geometry \phantom{\LARGE{A}} & $\vartheta$ & $\varsigma_{\mu\nu}$ & $\mathcal{R}$ \\ 
 \hline
 (ii) idealized cosmology \phantom{\LARGE{A}} & $\theta$ & $\sigma_{\mu\nu}$ & ${}^3\hspace{-0.05cm}R$\\
 \hline
 (iii) optical quantities \phantom{\LARGE{A}} & $\tilde{\theta}$ & $\tilde{\sigma}_{\mu\nu}$ & -- \\
 \hline
\end{tabular}
\end{center}
The reader may note that ``scalar curvature'' is not defined in the case of the optical quantities in the third row. They may also note that in the text we use $\tilde{\omega}_{\mu \nu}$ to denote the vorticity of null rays, and in Appendix A we use ${}^3\hspace{-0.05cm}R^*_{\mu \nu}$ to denote the trace-free part of the Ricci curvature tensor in the idealized cosmological models considered there. We use $u^{\mu}$ and $n^{\mu}$ to correspond to unit time-like vectors in the underlying space-time and the idealized cosmological models, respectively.


\begin{thebibliography}{}

\bibitem{many}
Verde, L., Sch$\ddot{\rm o}$neberg, N., \& Gil-Mar\'{i}n, H.,
A tale of many $ H_0$. 
{\it arXiv}:2311.13305.

\bibitem{cmb}
Aghanim, N. et al.,
Planck 2018 results-VI. Cosmological parameters. 
{\it Astronomy \& Astrophysics} 641, A6 (2020).

\bibitem{shoes}
Riess, A.~G et al., 
A comprehensive measurement of the local value of the Hubble constant with $1\, {\rm km} \, {\rm s}^{-1} \, {\rm Mpc}^{-1}$ uncertainty from the Hubble Space Telescope and the SH0ES team. 
{\it Astrophys. J. Lett.} 934, L7 (2022).

\bibitem{review}
Di Valentino, E., et al.,
In the realm of the Hubble tension—a review of solutions. 
{\it Class. Quant. Grav.} 38, 153001 (2021).

\bibitem{col1}
Collins, C.~B., \& Szafron, D.~A.,
A new approach to inhomogeneous cosmologies: Intrinsic symmetries. I. 
{\it J. Math. Phys.} 20, 2347 (1979).

\bibitem{col2}
Szafron, D.~A., \& Collins, C.~B.,
A new approach to inhomogeneous cosmologies: intrinsic symmetries. II. Conformally flat slices and an invariant classification. 
{\it J. Math. Phys.} 20, 2354 (1979).

\bibitem{col3}
Collins, C.~B., \& Szafron, D.~A.
A new approach to inhomogeneous cosmologies: Intrinsic symmetries. III. Conformally flat slices and their analysis.
{\it J. Math. Phys.} 20, 2362 (1979).

\bibitem{fitting}
Ellis, G.~F., \& Stoeger, W.,
The `fitting problem' in cosmology. 
{\it Class. Quant. Grav.} 4, 1697 (1987).

\bibitem{buchert1}
Buchert, T.,
On average properties of inhomogeneous fluids in general relativity: dust cosmologies. 
{\it Gen. Rel. Grav.} 32, 105 (2000).

\bibitem{buchert2}
Buchert, T.,
On average properties of inhomogeneous fluids in general relativity: perfect fluid cosmologies. 
{\it Gen. Rel. Grav.} 33, 1381 (2001).

\bibitem{ras1}
R$\ddot{\rm a}$s$\ddot{\rm a}$nen, S.,
Light propagation in statistically homogeneous and isotropic dust universes. 
{\it JCAP} 02, 011 (2009).

\bibitem{ras2}
R$\ddot{\rm a}$s$\ddot{\rm a}$nen, S.,
Light propagation in statistically homogeneous and isotropic universes with general matter content. 
{\it JCAP} 03, 018 (2010).

\bibitem{ten3}
Wagner, J.,
Solving the Hubble tension a la Ellis \& Stoeger 1987. 
{\it arXiv}:2203.11219.

\bibitem{rev}
Buchert, T. \& R$\ddot{\rm a}$s$\ddot{\rm a}$nen, S., 
Backreaction in late-time cosmology.
{\it Ann. Rev. Nuc. Part. Sci.} 62, 57 (2012).

\bibitem{lrs}
Ellis, G.~F.~R., 
Dynamics of pressure-free matter in general relativity.
{\it J. Math. Phys.} 8, 1171 (1967).

\bibitem{elst}
van~Elst, H. \& Ellis, G.~F.~R., 
The covariant approach to LRS perfect fluid spacetime geometries.
{\it Class. Quant. Grav.} 13, 1099 (1996).

\bibitem{wyman} 
Wyman, M., 
Non-static radially symmetric distributions of matter.
{\it Can. Math. Bull.} 19, 343 (1976).

\bibitem{exact}
Stephani, H., Kramer, D., MacCallum, M., Hoenselaers, C. \& Herlt, E., 
Exact solutions of Einstein's field equations. 
Cambridge University Press (2009).

\bibitem{sachs}
Sachs, R.,
Gravitational waves in general relativity. VI. The outgoing radiation condition. 
{\it Proc. Ro.l Soc. Lon. Ser. A. Math. Phys. Sci.} 264, 309 (1961).

\bibitem{eth}
Etherington, I.~M.~H., 
On the definition of distance in general relativity, 
{\it Phil. Mag.} 15, 761 (1933).

\bibitem{syksy2}
Lavinto, M., R$\ddot{\rm a}$s$\ddot{\rm a}$nen, S., \& Szybka, S. J.,
Average expansion rate and light propagation in a cosmological Tardis spacetime. 
{\it JCAP} 12, 051 (2013).

\bibitem{kok2}
Koksbang, S. M.,
Towards statistically homogeneous and isotropic perfect fluid universes with cosmic backreaction.
{\it Class. Quant. Grav.} 36, 185004 (2019).

\bibitem{kok3}
Koksbang, S. M.,
Another look at redshift drift and the backreaction conjecture. 
{\it JCAP} 10, 036 (2019).

\bibitem{kok4}
Koksbang, S. M.,
Observations in statistically homogeneous, locally inhomogeneous cosmological toy models without FLRW backgrounds. 
{\it MNRAS Lett.} 498, L135 (2020).

\bibitem{shear1}
Munshi, D., Valageas, P., Van~Waerbeke, L., \& Heavens, A., 
Cosmology with weak lensing surveys. 
{\it Phys. Rep.} 462, 67 (2008).

\bibitem{br1}
R$\ddot{\rm a}$s$\ddot{\rm a}$nen, S,
Evaluating backreaction with the peak model of structure formation. 
{\it JCAP} 04, 026 (2008).

\bibitem{br2}
Macpherson, H.~J., Lasky, P.~D., \& Price, D.~J.,
Inhomogeneous cosmology with numerical relativity. 
{\it Phys. Rev. D} 95, 064028 (2008).

\bibitem{br3}
Bolejko, K.,
Relativistic numerical cosmology with silent universes. 
{\it Class. Quant. Grav.} 35, 024003 (2017).

\bibitem{br4}
Adamek, J., Clarkson, C., Daverio, D., Durrer, R., \& Kunz, M.,
Safely smoothing spacetime: backreaction in relativistic cosmological simulations. 
{\it Class. Quant. Grav.} 36, 014001 (2018).

\bibitem{umeh}
Umeh, O.,
Consequences of using a smooth cosmic distance in a lumpy universe. I. 
{\it Phys. Rev. D} 106, 023514 (2022).

\bibitem{desi1}
Adame, A.~G., et al., 
DESI 2024 III: Baryon Acoustic Oscillations from Galaxies and Quasars. 
{\it arXiv}:2404.03000.

\bibitem{desi2}
Adame, A.~G., et al., 
DESI 2024 IV: Baryon Acoustic Oscillations from the Lyman Alpha Forest. 
{\it arXiv}:2404.03001.

\bibitem{sdss}
Alam, S., et al., 
Completed SDSS-IV extended Baryon Oscillation Spectroscopic Survey: Cosmological implications from two decades of spectroscopic surveys at the Apache Point Observatory. 
{\it Phys. Rev. D} 103, 083533 (2021).

\bibitem{sdss2}
Gil-Mar\'{i}n, et al.,
The Completed SDSS-IV extended Baryon Oscillation Spectroscopic Survey: measurement of the BAO and growth rate of structure of the luminous red galaxy sample from the anisotropic power spectrum between redshifts 0.6 and 1.0. 
{\it MNRAS} 498, 2492 (2020).

\bibitem{zal}
Coley, A.~A., Pelavas, N., \& Zalaletdinov, R.~M.,
Cosmological solutions in macroscopic gravity. 
{\it Phys. Rev. Lett.} 95, 151102 (2005).

\bibitem{sang1}
Sanghai, V.~A., \& Clifton, T.,
Post-Newtonian cosmological modelling. 
{\it Phys. Rev. D} 91, 103532 (2015).

\bibitem{sang2}
Sanghai, V.~A., \& Clifton, T.,
Cosmological backreaction in the presence of radiation and a cosmological constant. 
{\it Phys. Rev. D} 94, 023505 (2016).

\bibitem{desi3}
Adame, A.~G., et al.,
DESI 2024 VI: Cosmological Constraints from the Measurements of Baryon Acoustic Oscillations. 
{\it arXiv}:2404.03002.

\bibitem{ten1}
Heinesen, A., \& Buchert, T.,
Solving the curvature and Hubble parameter inconsistencies through structure formation-induced curvature. 
{\it Class. Quant. Grav.} 37, 164001 (2020).

\bibitem{temp1}
Larena, J., Alimi, J.~M., Buchert, T., Kunz, M., \& Corasaniti, P.~S.,
Testing backreaction effects with observations. 
{\it Phys. Rev. D} 79, 083011 (2009).

\bibitem{temp2}
Desgrange, C., Heinesen, A., \& Buchert, T.,
Dynamical spatial curvature as a fit to type Ia supernovae. 
{\it Int. J. Mod. Phys. D} 28, 1950143 (2019).

\bibitem{ten2}
Bolejko, K.,
Emerging spatial curvature can resolve the tension between high-redshift CMB and low-redshift distance ladder measurements of the Hubble constant. 
{\it Phys. Rev. D} 97, 103529 (2009).

\bibitem{inter}
Abdalla, E., et al.,
Cosmology intertwined: A review of the particle physics, astrophysics, and cosmology associated with the cosmological tensions and anomalies. 
{\it JHEP} 34, 49 (2022).

\bibitem{syksy3}
Boehm, C., \& R$\ddot{\rm a}$s$\ddot{\rm a}$nen, S.,
Violation of the FRW consistency condition as a signature of backreaction. 
{\it JCAP} 09, 003 (2013).

\bibitem{syksy4}
Montanari, F., \& R$\ddot{\rm a}$s$\ddot{\rm a}$nen, S.,
Backreaction and FRW consistency conditions. 
{\it JCAP} 11, 032 (2017).

\bibitem{kok1}
Koksbang, S. M., 
Searching for signals of inhomogeneity using multiple probes of the cosmic expansion rate H(z). 
{\it Phys. Rev. Lett.} 126, 231101 (2021).

\end{thebibliography}
\end{document}